\documentclass[twocolumn,aps]{revtex4}
\usepackage{graphicx}

\begin{document}
\title{Bell Measurement and Local
Measurement in the Modified Lo-Chau Quantum Key Distribution
Protocol}

\author{Won-Young Hwang}

\email{wyhwang@chonnam.ac.kr}
\affiliation{Department of Physics
Education, Chonnam National University, Kwangjoo 500-757, Republic
of Korea}

\begin{abstract}
We clarify the argument on the how (nonlocal) degenerate Bell
measurement can be replaced by local measurements in the modified
Lo-Chau quantum key distribution protocol. Discussing security
criterion for users, we describe how eavesdropper's refined
information on the final state is not helpful. We argue that current
discussions on the equivalence of the Bell and the local
measurements are not clear. We show how the problem of equivalence
can be resolved using the fact that eavesdropper's refined
information is not helpful for her.

\end{abstract}
\maketitle
\section{introduction}
Quantum cryptography, more precisely, quantum key distribution (QKD)
\cite{bene,maye,biha,shor,lo,gott,got2,gisi,brus,hwa2}, is one of
the most promising protocols in quantum information processing
\cite{niel}. Bennett-Brassard 1984 (BB84) QKD protocol \cite{bene}
had been widely conjectured to be secure based on the quantum
no-cloning theorem \cite{diek,woot,yuen} before its security was
shown by a few authors recently \cite{maye,biha,shor}. However,
Refs. 2 and 3 seem to be too complicated to be widely understood
while Ref. 4 is relatively simple in that it makes use of tools that
are familiar to quantum information scientist, e.g. quantum error
correcting code (QECC) and entanglement distillation protocol (EDP).
Thus, the approach of Shor and Preskill \cite{shor} is being more
widely accepted and applied to deal with the security of variations
of the BB 84 protocols \cite{brus,hwa2}.

In Ref. 4, the authors argue that the modified Lo-Chau (LC) protocol
based on the EDP is secure. Then, they show that the modified LC
protocol reduces to the BB84 protocol. Thus, the security of the
BB84 protocol depends on that of the modified LC protocol. In the
discussion on security of the modified LC protocol, they use
`classicalization of statistics' (or `quantum to classical
reduction') \cite{lo,gott}. However, in the derivation of the
classicalization of statistics, they make use of a (partial)
equivalence between (degenerate) Bell measurements and local
measurements: Bell measurements can be replaced by $Z$ measurement
in the $|0\rangle, |1\rangle$ basis and by $X$ measurement in the
$|\bar{0}\rangle, |\bar{1}\rangle$ basis. Here, a Bell measurement
is one in the Bell basis $ |\Phi^{\pm}\rangle=
(1/\sqrt{2})(|0\rangle_A |0\rangle_B \pm |1\rangle_A |1\rangle_B)$
 and
$ |\Psi^{\pm}\rangle= (1/\sqrt{2})(|0\rangle_A |1\rangle_B \pm
|1\rangle_A |0\rangle_B)$, where $A$ and $B$ denote two users, Alice
and Bob, respectively.
 $|\bar{0}\rangle= (1/\sqrt{2})(|0\rangle+ |1\rangle)$ and
        $ |\bar{1}\rangle= (1/\sqrt{2})(|0\rangle- |1\rangle)$.
(The $Y$ measurment is the one in the basis $|\tilde{0}\rangle=
(1/\sqrt{2})(|0\rangle +i |1\rangle)$ and $|\tilde{1}\rangle=
(1/\sqrt{2})(|0\rangle -i |1\rangle)$.) However, the discussion on
the equivalence is not clear as we will see. On the other hand, the
security considered so far has been that from Eve's (eveasdropper's)
point of view. However, what we eventually need is a security
criterion for Alice and Bob, the users.

The purpose of this paper is to give a clearer presentation for the
equivalence between the Bell and the local measurements in the
modified LC protocol. While discussing security criteria for Alice
and Bob, we show that Eve's refined information on the final state
is not helpful to her. Then, we clarify the equivalence between Bell
measurements and $Z$ and $X$ measurements.

This paper is organized as follows: First, we reformulate the
modified LC protocol and arguments for its security in an explicit
manner. Next, briefly discussing security criteria for Alice and
Bob, we show that Eve's refined information on the final state is
not helpful to her. Then, we argue that the current discussions on
the equivalence of the Bell and local measurements are not clear,
and we show that the problem of the equivalence can be resolved
using the fact that Eve's refined information is not helpful to her.
\section{Modified Lo-Chau protocol}
The legitimate states in the modified LC protocol are pairs of a
Bell state, $|\Phi^+\rangle$. However, the whole quantum state that
Alice, Bob, and Eve share, after distribution of quantum bits
(qubits)
 and before entanglement distillation, is an arbitrary
state $|\psi_{ABE}\rangle$ that Eve chooses. Let us write down the
state  $|\psi_{ABE}\rangle$ in the Bell basis:
\begin{equation}
\label{a}
 |\psi_{ABE}\rangle =
\sum_{\{k\}} C_{\{k\}}
\sigma_{{k_1}(1)} \sigma_{{k_2}(2)} \cdot \cdot \cdot
 \sigma_{{k_{2n}}(2n)}
|\Phi^+\rangle^{\otimes 2n} |E_{\{k\}} \rangle.
\end{equation}
Here, $\{k\}$ is an abbreviation for $k_1,k_2,...,k_{2n}$ with
 $k_i= 0,1,2,3$ ($i=1,2,...,2n$), and $\sigma_0= I$,
 $\sigma_1= X$, $\sigma_2= Y$,
 $\sigma_3= Z$ are Pauli operators.
The $\sigma_{k_i(i)}$ denotes the Pauli operator $\sigma_{k_i}$
acting on $i$th qubit of Bob. Namely, $\sigma_{k_i(i)}$ is
$I \otimes \sigma_{k_i} $ acting on $i$th qubit pairs that are shared
by Alice and Bob. Note that the set of
 $\sigma_{{k_1}(1)} \sigma_{{k_2}(2)} \cdot \cdot \cdot
 \sigma_{{k_{2n}}(2n)}  |\Phi^+\rangle^{\otimes 2n}$ of all $\{k\}$
 constitutes the complete Bell basis for the $2n$ qubit pairs.
The $C_{\{k\}}$'s are coefficients in complex numbers. Eve's states
$|E_{\{k\}} \rangle$ are normalized, but not mutually orthogonal in
general. It is notable that the state in Eq. (\ref{a}) is completely
general; thus, it is dealing with all attacks including opaque
(intercept-resend), individual, collective, and joint attacks
\cite{gisi}.

Let us describe the checking method.
Consider a measurement $M_Z$
 whose projection operators are
\begin{eqnarray}
\label{b}
P_0 &=& |\Phi^+\rangle \langle \Phi^+| +|\Phi^-\rangle \langle \Phi^-|
= |00\rangle \langle 00| +|11\rangle \langle 11|,
\nonumber\\
P_1 &=& |\Psi^+\rangle \langle \Psi^+| +|\Psi^-\rangle \langle \Psi^-|
= |01\rangle \langle 01| +|10\rangle \langle 10|.
\end{eqnarray}
Also, consider a measurement $M_X$ whose projection operators are
\begin{eqnarray}
\label{c}
\bar{P}_0 &=&
 |\Phi^+\rangle \langle \Phi^+| +|\Psi^+\rangle \langle \Psi^+|
= |\bar{0}\bar{0}\rangle \langle \bar{0} \bar{0}|
+|\bar{1} \bar{1}\rangle \langle \bar{1} \bar{1}|,
\nonumber\\
\bar{P}_1 &=& |\Phi^-\rangle \langle \Phi^-|
+|\Psi^-\rangle \langle \Psi^-|
= |\bar{0}\bar{1}\rangle \langle \bar{0} \bar{1}|
+|\bar{1} \bar{0}\rangle \langle \bar{1} \bar{0}|.
\end{eqnarray}
The measurements $M_Z$ and $M_X$ are {\it nonlocal}. (For example,
$\bar{P}_0 |00\rangle = (1/\sqrt{2}) \bar{P}_0 (|\Phi^+\rangle+
|\Phi^-\rangle)= (1/\sqrt{2}) |\Phi^+\rangle$; that is, a separable
state $|00\rangle$ is transformed to a nonlocal state $
|\Phi^+\rangle$.) Thus, obviously they cannot be performed by Alice
and Bob who are supposedly separated. However, we assume that they
can perform the measurements $M_Z$ and $M_X$ for the time being. We
will see later how $M_Z$ and $M_X$ can be replaced by separable
measurements $Z$ and $X$.

The error rate in the modified LC protocol is defined as follows:
First Alice and Bob randomly choose $n$ pairs of qubits from among
the $2n$ pairs. On each of the $n$ chosen pairs, they perform a
measurement randomly chosen between $M_Z$ and $M_X$.
%
The error rate $e$ is the number of all instances when the
measurement outcomes are those corresponding to $P_1$ and
$\bar{P}_1$ divided by the number of samples $n$. From Eqs.
(\ref{b}) and (\ref{c}), we can see that the legitimate state
$|\Phi^+\rangle$ has zero probability to give rise to an error, and
that other illegitimate states, $|\Phi^-\rangle$, $|\Psi^+\rangle$,
and $|\Psi^-\rangle$, have non-zero probabilities, $1/2, 1/2,$ and
$1$, respectively, to give rise to an error.

Let us now describe the `classicalization of statistics'
\cite{lo,gott}. We consider a case $n= 2$, which is simple but
illustrative enough. Assume that Alice and Bob's random choice was
to measure the first and the third pairs in the $Z$ and the $X$
bases, respectively. That is, they perform a measurement $M_Z
\otimes I \otimes M_X \otimes I $ on a state
$|\psi^{\prime}_{ABE}\rangle =
\sum_{\{k\}} C_{\{k\}}
\sigma_{{k_1}(1)} \sigma_{{k_2}(2)}
\sigma_{{k_3}(3)} \sigma_{{k_4}(4)}
|\Phi^+\rangle^{\otimes 4} |E_{\{k\}} \rangle$.
It is easy to see that, for example,
the probability $p_{00}$
 that they get $0$ in the $M_Z$ measurement
and $0$ in the $M_X$ measurement is given by $ p_{00} =
\sum_{k_1=0,3, k_2, k_3= 0,1, k_4} |C_{\{k\}}|^2$, where
 $k_2, k_4 = 0,1,2,3$.
The resultant state is
$ |\psi_{00}^{\prime}\rangle = N
 \sum_{k_1=0,3, k_2, k_3=0,1, k_4} $ $
 C_{\{k\}} \sigma_{{k_1}(1)} \sigma_{{k_2}(2)}
\sigma_{{k_3}(3)} \sigma_{{k_4}(4)}
|\Phi^+\rangle^{\otimes 4} |E_{\{k\}} \rangle$, where
$N=1/\sum_{k_1=0,3, k_2, k_3=0,1, k_4}
 |C_{\{k\}}|^2 $ is the normalization constant.
(Note Eqs. (\ref{b}) and (\ref{c}) and
 that $(I \otimes \sigma_3) |\Phi^+\rangle= |\Phi^-\rangle $
and  $(I \otimes \sigma_1) |\Phi^+\rangle= |\Psi^+\rangle$.) In the
same way, we can calculate the probabilities $p_{01}$, $p_{10}$, and
$p_{11}$ and the corresponding resultant states. Then, let us
consider a case where Eve prepares a mixed state
$\rho^{\prime}= \sum_{\{k\}} P_{\{k\}}
\sigma_{{k_1}(1)} \sigma_{{k_2}(2)}
\sigma_{{k_3}(3)} \sigma_{{k_4}(4)}
|\Phi^+\rangle ^{\otimes 4}
\\
 \langle \Phi^+|^{\otimes 4}
\sigma_{{k_1}(1)} \sigma_{{k_2}(2)} \sigma_{{k_3}(3)}
\sigma_{{k_4}(4)}$,
where $ P_{\{k\}} \equiv |C_{\{k\}}|^2 $.
We consider the case where
 Alice and Bob perform the same checking measurement
$M_Z \otimes I \otimes M_X \otimes I $ on the state $\rho^{\prime}$.
Then, it is easy to see that, for example, the probability $q_{00}$
that they get $0$ in the $M_Z$ measurement and $0$ in the $M_X$
measurement is the same as the $p_{00}$ given above. However, the
resultant state is not the same, but is given by $
\rho_{00}^{\prime} = N
 \sum_{k_1=0,3, k_2, k_3=0,1, k_4} $ $
 P_{\{k\}} \sigma_{{k_1}(1)} \sigma_{{k_2}(2)}
\sigma_{{k_3}(3)} \sigma_{{k_4}(4)}
|\Phi^+\rangle^{\otimes 4}$ $
 \langle \Phi^+|^{\otimes 4}
\sigma_{{k_1}(1)} \sigma_{{k_2}(2)} \sigma_{{k_3}(3)}
\sigma_{{k_4}(4)} $. Note a similarity between the states $
|\psi_{00}^{\prime}\rangle $ and $\rho_{00}^{\prime}$. The
difference is that the former is full superpositions that are
partially broken in the latter. In both cases, the only states that
are compatible with the measurement outcomes remain with the
relative magnitude of $C_{\{k\}}$'s and $ P_{\{k\}}$'s preserved. In
the same way, we can calculate $p^{\prime}_{01}$,
 $ p^{\prime}_{10}$, and $p^{\prime}_{11}$  that also turn out to be
 the same as
$p_{01}$, $ p_{10}$, and $ p_{11}$, respectively, and we obtain the
corresponding resultant states. It is straightforward to generalize
the calculations to the general $n$ case. If the similarity in the
resultant states is used, it is not difficult to see the
classicalization of statistics \cite{lo,gott}: For each state
$|\psi_{ABE}\rangle$ in Eq. (\ref{a}),
 there exists a corresponding state
\begin{eqnarray}
 \label{eb}
\rho = \sum_{\{k\}} P_{\{k\}} &&
\sigma_{{k_1}(1)}
\sigma_{{k_2}(2)} \cdot \cdot \cdot \sigma_{{k_{2n}}(2n)}
|\Phi^+\rangle^{\otimes 2n} \nonumber\\
&& \langle \Phi^+|^{\otimes 2n}
\sigma_{{k_1}(1)}
\sigma_{{k_2}(2)} \cdot \cdot \cdot \sigma_{{k_{2n}}(2n)}
\end{eqnarray}
that has the following properties: First, the state gives rise to
the same statistics for the checking measurement and, thus, has the
same probability to pass the test as those of the state $
|\psi_{ABE}\rangle$. Second, when it passes the test, the resultant
states give rise to, after the EDP, a state that has the same
fidelity as the legitimate states, as in the case of the state
$|\psi_{ABE}\rangle$. It is the fidelity of the final state
$S(\rho_{AB})$ to the legitimate state that bounds Eve's information
on the final key, as seen below.
 Thus as far as the security is concerned, the two
states in Eqs. (\ref{a}) and (\ref{eb}) are equivalent. The
classicalization of statistics plays a crucial role in showing the
security in that it removes some entanglement in the state of Eq.
(\ref{a}) that is otherwise intractable for Alice and Bob.

Now let us see how the fidelity of the final state of Alice and Bob
to the legitimate state bounds Eve's information on the final key
\cite{lo,gott}. Eve knows the initial state that she prepared, and
she knows the identities of all quantum processing that has been
done on the initial state, including identities and outcomes of
measurements, and a chosen QECC because they are publicly announced.
 Therefore, Eve can calculate the
final (pure) state $\rho_{ABE}$ that Alice, Bob, and herself share.
Alice and Bob's state $\rho_{AB}$ and Eve's state $\rho_{E}$ are
given by $ \rho_{AB} = \mbox{tr}_E (\rho_{ABE}) $ and $ \rho_{E} =
\mbox{tr}_{AB} (\rho_{ABE})$, respectively. Eve's von Neumann
entropy $S(\rho_E) = -\mbox{tr}(\rho_E \log \rho_E)$ is the same as
Alice and Bob's $S(\rho_{AB})$ because $\rho_{ABE}$ is a pure state.
On the other hand, it can be observed \cite{lo,gott} that the mutual
information between Eve's party and Alice and Bob's party,
$I(AB;E)$, is bounded by Holevo's theorem \cite{niel}
\begin{equation}
\label{g}
 I(AB;E) \leq  S(\rho_{E}) =S(\rho_{AB}).
\end{equation}
However, $S(\rho_{AB})$ is bounded by the fidelity of the state
$\rho_{AB}$ to the legitimate state \cite{lo,gott}.

Let us discuss the arguments for security of the modified LC
protocol. Assume that Eve distributed the state in Eq. (\ref{a}) to
Alice and Bob. Due to the classicalization of statistics, however,
it is sufficient for them to consider a corresponding case where Eve
distributed the state in Eq. (\ref{eb}). Thus, first we may well
separately consider each term
\begin{eqnarray}
\label{ec}
\sigma_{{k_1}(1)}&&
\sigma_{{k_2}(2)} \cdot \cdot \cdot \sigma_{{k_{2n}}(2n)}
|\Phi^+\rangle^{\otimes 2n}
\nonumber\\
&&\langle \Phi^+|^{\otimes 2n}
\sigma_{{k_1}(1)}
\sigma_{{k_2}(2)} \cdot \cdot \cdot \sigma_{{k_{2n}}(2n)}
\end{eqnarray}
with a particular $\{k\}$  and then combine them later. Consider the
checking measurement where $n$ randomly chosen pairs are measured
along randomly chosen bases. We can see that the error rate $e$ is
statistically proportional to
 the ratio of the illegitimate states among the
 $n$ checked pairs.
Alice and Bob abort the protocol if the measured error rate $e$ is
larger than a threshold for checking, $e_{check}$. The threshold for
checking, $e_{check}$, is set to be a little bit smaller than a
threshold for error correction $e_{cor.}$, to compensate for
statistical fluctuations. Let us assume that the number of
illegitimate states in the state in Eq. (\ref{ec}) is larger than
$(2n) (2 e_{cor.})$. Since the probability that each illegitimate
state is detected in the checking procedure is equal to or larger
than $1/2$, as seen above,
 it typically give rise to
an error rate larger than $e_{cor.}$ for the checked pairs. Then,
the probability that the state passes the test is negligibly small.
In other words, the checking procedure sifts out, with high
probability, any state that contains more than $4 n e_{cor.} $ pairs
of illegitimate states. Combining this fact with Bayes's theorem, we
get the following: Whatever state in Eq. (\ref{eb}) Eve has
prepared, if the state passes the test with a non-negligible
probability, then the ratio of illegitimate states of the resultant
state is less than $2e_{cor.}$ with high probability. Therefore, if
they choose a QECC that can correct up to an error rate $2e_{cor.}$,
the final state after quantum error correction by using the chosen
code will have a high fidelity to the legitimate state.
\section{Security criterion and Eve's refined information}
Let us now discuss the security criterion. What has been considered
so far is Eve's viewpoint: The higher the probability to pass the
test by Alice and Bob, the less the information that Eve gets is.
However, what we actually need is the security criterion for Alice
and Bob, the users. It is intuitively clear that the security
criterion for Eve can be translated to that for Alice and Bob.
However, there remain a few difficulties in doing so because Alice
and Bob do not know the initial state that Eve knows. Eve's most
general strategy is to prepare a state in Eq. (\ref{eb}) with a
probability distribution $P_{\{k\}}$'s. Once Alice and Bob know the
probability distribution $P_{\{k\}}$'s, combined with the Bayes's
theorem, they can calculate the final state. However, the problem is
that it is not clear what probability distribution $P_{\{k\}}$'s Eve
will choose. However, we can say that Eve will optimize her
strategy. That is, she will choose an attack that maximizes
 her information on the key among those with
 the same probability to pass the test.
Identification of the optimal probability distribution $P_{\{k\}}$'s
seems to be at heart of the open problem, to find a clear security
criterion for Alice and Bob \cite{gott}.

However, a certain probability distribution $P_{\{k\}}$'s has been
tacitly assumed in discussions on QKD so far. For example, let us
consider the case where Eve chooses an attack that can give her full
information on the key once it pass the test, but the probability to
pass the test is negligible.
 That is, she chooses
that $P_{\{k\}}\neq 0$ for only those $\{k\}$'s in which most of
 $k_i$ are
nonzero. If Alice and Bob assume that Eve adopts this strategy,
combined with the Bayes's theorem, their conclusion is always that
Eve has full information. However, this is not regarded as a
strategy that Eve will actually adopt because it blocks
communication between Alice and Bob. Here we accept that it is not
 a good strategy for Eve.
 Here, we do not try to get a rigorous security criterion.
It may be a very subtle problem to find Eve's optimal strategy
because it may depend on the real situations in which Alice, Bob,
and Eve find themselves. However, it is reasonable to say that Eve's
optimal strategy is such that Alice and Bob get a state $\rho_{AB}$
that is almost a legitimate state as a result of the EDP, as assumed
so far.

The situation we meet here is that  Alice and Bob have only
 partial information on the state for which
 Eve has full information.
That is, we have
\begin{equation}
\label{f}
\rho_{AB}= \sum_i p_i \mbox{tr}_E (\rho_{ABE}^i)
          \equiv  \sum_i p_i \rho_{AB}^i.
\end{equation}
Here, $\rho_{ABE}^i$ denotes a (pure) state inferred from Eve's
 classical information $i$.
Consider a case where $F(\rho_{AB}, |\Phi \rangle ^{\otimes k}) \geq
1-\epsilon$ with $F$ denoting the fidelity \cite{niel} and
$\epsilon$ being a positive real number.
 Let us see how
Eve's  refined knowledge on the state is not so helpful to her by
using the following two arguments.

Let us give the first argument. For each $\rho_{ABE}^i$, Eve's state
is given by $ \rho_E^i = \mbox{tr}_{AB} (\rho_{ABE}^i)$, and
$S(\rho_E^i) = S(\rho_{AB}^i)$ because $\rho_{ABE}^i$ is a pure
state. Then, the mutual information between Eve's party and Alice
and Bob's party, $I^i(AB;E)$, is bounded by the $S(\rho_{AB}^i)$.
However,  we can see that
\begin{equation}
\label{h} 1-\epsilon \leq F(\rho_{AB}, |\Phi \rangle ^{\otimes k}) =
\sum_i p_i F(\rho_{AB}^i, |\Phi \rangle ^{\otimes k})
\end{equation}
by using Eq. (\ref{f}) and the relation $F(\sum_i p_i \rho^i,
|\psi\rangle)= \sum_i p_i F(\rho^i, |\psi\rangle)$.
Equation (\ref{h}) says that the average of the fidelities of
$\rho_{AB}^i$ is bounded by a quantity $1-\epsilon$ that bounds the
fidelity of the average (mixed) state $\rho_{AB}$. Here, Eve cannot
control the outcome of the classical information $i$. Thus, it is a
meaningful average even if Eve knows the $i$. More concretely, let
us consider a particular $F(\rho_{AB}^i, |\Phi \rangle ^{\otimes
k})$. From Eq. (\ref{h}), we have that $\{1- F(\rho_{AB}^i, |\Phi
\rangle ^{\otimes k})\} p_i \geq \epsilon$, which expresses a
reciprocal relation between the closeness of the state to legitimate
states and a probability that it happens.

Let us give the second argument.
Consider a purification of the state $\rho_{AB}$
\begin{equation}
\label{i} |\psi_{ABE}\rangle= \sum_i \sqrt{p_i} |i\rangle
|\psi^i_{ABE}\rangle
\end{equation}
that is compatible with Eq. (\ref{f}). Here, $ |\psi^i_{ABE}\rangle
\langle \psi^i_{ABE}| = \rho^i_{ABE}$, and the qubits storing the
classical information $i$ are at Eve's hands. If Eve first performs
a measurement on qubits storing the $i$, then the state reduces to
$\rho^i_{ABE}$ with probability $p_i$. The situation now is
equivalent to the case above where Eve has the state $\rho^i_{ABE}$
with knowledge of the $i$.
However, in this case the bound in Eq. (\ref{g}) is valid for the
pure state $|\psi_{ABE}\rangle$.
 Therefore, we can say that the bound in Eq.
(\ref{g}) applies to the case where Eve has more refined information
$i$ about the final state.
\section{Equivalence of the Bell and the local measurements}
Now let us discuss the problem of partial equivalence. So far, we
have assumed that Alice and Bob can perform $M_Z$ and  $M_X$.
However, they are nonlocal measurements that cannot be actually done
by Alice and Bob, as seen above. Their argument for this problem is
the following \cite{gott,got2,lo,shor}: Let us consider the actual
situation where Alice and Bob each perform $Z$ measurements on a
pair of qubits. The basis for this measurement is $|00\rangle$,
$|01\rangle$, $|10\rangle$, and $|11\rangle$.
By Eq. (\ref{b}), however, we can estimate the probabilities
involved with the $M_Z$ measurement solely from the outcomes of the
$Z$ measurement: For example, $\mbox{tr}\{\rho (|\Phi^+\rangle
\langle \Phi^+| +|\Phi^-\rangle \langle \Phi^-|) \} =
\mbox{tr}\{\rho (|00\rangle \langle 00| +|11\rangle \langle 11|) \}
= \mbox{tr}(\rho |00\rangle \langle 00|)+
   \mbox{tr}(\rho |11\rangle \langle 11|)$.
The same thing can be said for the $M_X$ and the $X$ measurements.
However, the measured (or checked) $n$ pairs are not further used by
Alice and Bob. Only the remaining $n$ information pairs are used for
key generation. That is, the checked and the information pairs are
different. Therefore, the density operator for the information pairs
is invariant to whatever measurement they do on the checked pairs,
provided that they do not make use of the information on their
measurements outcomes. (One's quantum state can depend on his/her
knowledge about the measurement outcomes on the other side when the
shared initial state is entangled, as is well known.) Alice and Bob
may not make use of the information on the measurement outcomes.
Therefore, if the error rate $e_L$ estimated by the local
measurements $Z$ and $X$ is below the threshold for checking
$e_{check}$,the  EDP will be successful with high probability: The
fact that $e_L$ is less than $e_{check}$ implies that if they had
estimated the error rate $e_B$ by performing the nonlocal
measurement $M_Z$ and $M_X$, then they also would have found that
$e_B$ was less than $e_{check}$ with high probability. However, as
we have seen above, if  $e_B$ is less than $e_{check}$, then the EDP
is successful with high probability.

However,  the measurement outcomes are publicly announced in the
protocol; thus, Eve knows it. {\it The problem is that Eve  can make
use of the information on the measurement outcomes.} Our solution to
this problem is the following: The best thing that Eve can get is
the refined knowledge on the final state, however, and we have seen
above that such refined information is not so helpful to Eve.
Therefore, now we can safely
 say that the local $Z$ and $X$ measurements are equivalent to
 the nonlocal $M_Z$ and $M_X$ measurements as far as the
 security of BB84 protocol is concerned.


\section{Conclusion}
To summarize, we reformulated the modified LC protocol and arguments
for its security in an explicit manner. While discussing the
security criterion for Alice and Bob briefly, we showed that Eve's
refined information on the final state is not helpful to her. Then,
we argued that current discussions on the equivalence of the Bell
and the local measurements are not clear, and we showed that the
problem of the equivalence can be resolved using the fact that Eve's
refined information is not helpful to her.

\acknowledgments I am grateful to Profs. Hoi-Kwong Lo, G. M. D'
Ariano, and Maxim Raginsky for helpful discussions.



\begin{references}
\bibitem{bene} C. H. Bennett and G. Brassard, Proc. IEEE
                Int. Conf. on Computers, Systems, and Signal
                Processing, Bangalore (IEEE, New York, 1984),
                p. 175.
\bibitem{maye} D. Mayers,
                 J. Assoc. Comput. Mach. {\bf 48}, 351 (2001);
                 quant-ph/9802025.
\bibitem{biha} E. Biham, M. Boyer, P. O. Boykin, T. Mor, and
               V. Roychowdhury, in {\it Proceedings of the
               Thirty-Second Annual ACM Symposium on Theory
               of Computing} (ACM Press, New York, 2000), p.
               715; quant-ph/9912053.
\bibitem{shor} P. W. Shor and J. Preskill, Phys. Rev. Lett.
               {\bf 85}, 441 (2000).
\bibitem{lo}  H-K. Lo and H. F. Chau, Science {\bf283},
              2050 (1999).
\bibitem{gott} D. Gottesman and H-K. Lo, IEEE Trans. Inform. Theory
               {\bf 49}, 457 (2003).
\bibitem{got2} D. Gottesman and J. Preskill, Phys. Rev. A {\bf 63},
                022309 (2001).
\bibitem{gisi} N. Gisin, G. Ribordy, W. Tittel, and H. Zbinden,
               Rev. Mod. Phys. {\bf 74}, 145 (2002).
\bibitem{brus} D. Bru\ss, Phys. Rev. Lett. {\bf81}, 3018 (1998);
                 H-K. Lo, Quan. Inf. Com. {\bf 1}, 81 (2001).
\bibitem{hwa2} W-Y. Hwang, I. G. Koh, and Y. D. Han,
                 Phys. Lett. A   {\bf 244}, 489 (1998);
                 W-Y. Hwang, X-B. Wang, K. Matsumoto, J. Kim, and
                 H-W. Lee, Phys. Rev. A {\bf 67}, 012302 (2003).
\bibitem{niel} M. A. Nielsen and I. L. Chuang, {\it Quantum Computation
                and Quantum Information} (Cambridge
                Univ. Press, Cambridge, U.K., 2000).
\bibitem{hwa2} W-Y. Hwang, J. Lee, D. Ahn, S. W. Hwang,
               J. Korean Phys. Soc. {\bf 42}, 167 (2003).
\bibitem{diek} D. Dieks, Phys. Lett. A {\bf92}, 271 (1982).
\bibitem{woot} W. K. Wootters and W. H. Zurek, Nature  {\bf299},
                802  (1982).
\bibitem{yuen} H. P. Yuen, Phys. Lett. A {\bf 113}, 405 (1986).
\end{references}
\end{document}